\documentstyle[twoside,fleqn,espcrc2,psfig]{article}

\def\mus{\mu {\rm sec}}
\def\esc2{{\rm erg} \, {\rm sec}^{-1} \, {\rm cm}^{-2}}
\def\es{{\rm erg} \, {\rm sec}^{-1} }

\newcommand{\AmS}{{\protect\the\textfont2
  A\kern-.1667em\lower.5ex\hbox{M}\kern-.125emS}}

\title{Peak Luminosities of Bursts from GRO J1744-28 measured with the 
	RXTE PCA; \\ {\it Italia:  wij post 17 two 4s - a one man marching band -
	got darn bright.}}

\author{Keith Jahoda, Michael J. Stark, Tod E. Strohmayer, William Zhang,
\address{Laboratory for High Energy Astrophysics,
	Goddard Space Flight Center, \\
	Code 662, Greenbelt, MD, 20771   USA}
Edward H. Morgan, and Derek Fox
\address{Center for Space Research, MIT, Cambridge, MA, 02138}
 }
       
\begin{document}

\begin{abstract}

GRO J1744-28, discovered by BATSE in December 95, is the second neutron star
system known to produce frequent accretion powered bursts.  The system has
been regularly monitored with the RXTE PCA since the peak of the first outburst
in January 96 at which time the observed persistent and bursting count rates
were $\sim$ 25,000 ct/sec and $\sim$ 150,000 ct/sec, with corresponding instrumental
deadtimes of $\sim 10\%$ and $\ge 50\%$.  
We present a model
which allows the reconstruction of the true incident count rate in the
presence of enormous deadtime and use the model to
derive a history of the peak luminosities
and fluences
of the bursts as a function of time.
During the peak of the January 1996 and January 1997
outbursts, when the 
persistent emission was $\ge 1$ Crab, we infer peak luminosities of $\sim 100$
times the
Eddington luminosity, and a ratio of persistent emission to integrated burst
emission of $\sim 34$.  

\end{abstract}

\maketitle

\section{GRO J1744-28}

GRO J1744-28, the ``bursting pulsar" was dicovered by the Burst And
Transient Source Experiment (BATSE) in December 1995 \cite{Kouv96}
just prior to the launch of the Rossi X-ray Timing Explorer (RXTE)
on 30 December 1995.  RXTE has  performed a series of observations
of this object from January 1996 through November 1997 during which
time the source has varied from having a persistent flux in excess
of 1 Crab with bursts which saturated the Proportional Counter Array
(PCA) to being undetectable.  The PCA observations complement the
nearly continuous BATSE record by virtue of the large collecting area
and high time resolution of the PCA and High Energy X-ray Timing
Experiment (HEXTE) experiments.  These pointed experiments can study
individual bursts in much greater detail and follow the evolution of 
the outbursts to much fainter fluxes than BATSE.  Studies of the
individual bursts has, until now, been handicapped by an incomplete
understanding of the deadtime processes in the PCA when observing extremely
bright sources.  The three main purposes of this contribution are to
(a)  present a model
of deadtime processes in the PCA which is reliable for input fluxes
at least up to 35 Crab (
$\sim 0.5 \times 10^6 $ count/sec);  
(b) establish that the  bursts from J1744-28 reached luminosities
of $\sim 100$ times the Eddington luminosity;  and
(c) to provide advice to users of RXTE who may observe sources
brighter than 5 Crab during the rest of the RXTE mission.

%

\section{EVIDENCE OF SATURATION}

Figure 1 shows, for one burst observed on 1996 February 4 and which peaked at
13:15:33 UTC,
the observed good event rate as well as the {\it Remaining Counts} rate
from the Standard 1 data plotted with 0.125 second time bins.  The Standard
1 data (present for all PCA observations) contains 8 rates, read out every
0.125 seconds and internal calibration spectra.  The eight rates consist
of the good rate in each of the 5 detectors, the sum over 5 detectors of all
events which trigger only the propane layer, all events which trigger the
VLE flag (i.e. saturating events), and all other events commonly called the
{\it Remaining Counts} rate.  For most observations, the remaining Counts
rate is dominated by particle induced background events;  for bright sources
there is a non trivial possibility that two (or more) cosmic X-rays will be
detected in different parts of the detector and be recorded as a multiple
anode event.  The sudden rise in the Remaining Counts rate, coincident in time
with the rise in the Good event rate, makes it clear that in the burst, the
remaining counts rate is dominated by the coincidences of photons which arise
in the burst.

It is however obvious that while the good event rate reaches an apparent
plateau around $1.5 \times 10^5$ count/sec, the remaining count rate has a 
structure with greater contrast.  This observation suggests the central
theme of our deadtime correction:  the rate of coincident events, which 
depends on the square of the incident rate, is a better measure of the 
flux of very bright sources than the good rate.  Information on the
coincidence rate is always available on 0.125 second scales, and can be available
with higher time resolution through the use of EDS modes which count and
telemeter events which trigger exactly two anode chains.  (For a complete list of 
modes see \cite{modes}.)

\begin{figure}
\psfig{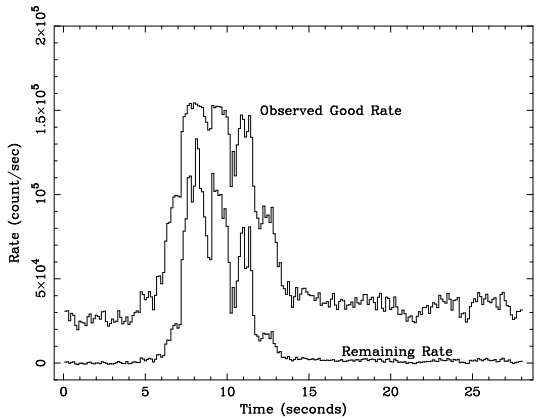}
\caption{Good and Coincidence rates observed from a 1996 February 4 burst.}
\label{fig1}
\end{figure}

Examination of numerous bursts from January and February 1996 show similar
characteristics:  the good rate saturates near $1.5 \times 10^5$ count/sec
while the remaining rate shows much larger relative variations.

\section{PCA AND EDS CHARACTERISTICS}

\subsection{PCA}

Each Proportional Counter Unit (PCU) has 9 independent signal chains.
Seven signal chains, designated L1, R1, L2, R2, L3, R3, and VP, signifying
the ``left" and ``right" halves of the three Xenon layers and the propane
veto layer \cite{Zhang95} have individual charge sensitive pre-amplifiers
and shaping amplifiers and share a common analog to digital converter (ADC).
The xenon veto anodes and the calibration flag set discriminators but are not
pulse height analyzed.  A good event is one which triggers exactly one
signal chain;  the pulse height from the ADC is unambiguously associated with
that signal chain.  An event which triggers two or more chains produces
two flags and a pulse height which could correspond to either chain (but in
any case which contains only a fraction of the deposited energy) and is generally
discarded as a non X-ray event.  (An event which triggers the calibration 
flag and one other anode chain is an exception to this rule and is tagged as
a calibration event.)  The analog pulse shaping is a paralyzable process
while the analog to digital conversion is a non-paralyzable process.

\subsection{EDS}

The EDS always runs two Standard modes in addition to the data modes
selected by the observer.  A key feature of each of these modes is that every
PCA event is recorded exactly once.  Standard 1 produces
8 rates read out every 0.125 seconds while Standard 2 produces a spectrum
for each signal chain and 29 distinct coincidence rates for each PCU
once every 16 seconds.  In special circumstances, the Standard 2 mode
can be read out on shorter time scales.  We present, in the next section,
data obtained while observing bright sources with the Standard 2 mode
being read out more frequently than usual.  These data allow us to 
parameterize a model of PCA deadtime processes.

\section{DEADTIME MODEL}

Here we present data obtained while slowly scanning over Sco X-1 (with a 1 second readout
interval for the Standard 2 data) and data obtained during a bright
burst from J1744-28 (with 2 second readout).   The details of the paralyzable
deadtime process associated with the analog pulse shaping are pulse height
dependent, so we expect some differences in parameters derived from Sco X-1
(a very soft source) and J1744-28 (a very hard source).

\begin{figure}
\psfig{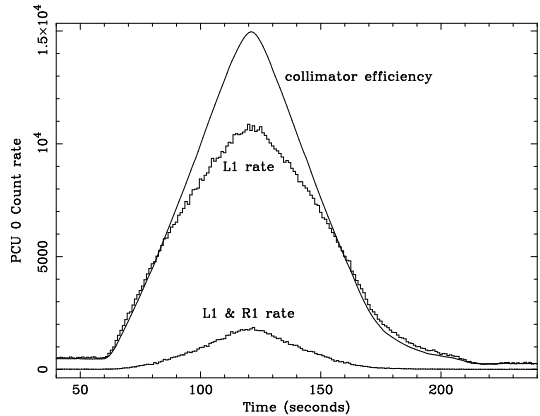}
\caption{Collimator transmission, L1 rate, and L1+R1 rate for one PCU during
scan over Sco X-1 in normal branch}
\label{fig3}
\end{figure}

Figure \ref{fig3} shows two rates observed during a scan over Sco X-1 (97Mar15).
The smooth curve represents the collimator efficiency for one of the PCU, and the
histograms represent the counting rate observed in the L1 layer (i.e. good
events) and the rate of L1 plus R1 coincidences (two photons observed within
the coincidence window, one on each half of the first layer).  The collimator efficiency is
scaled to demonstrate that  the L1 rate is affected by a $25-30\%$ deadtime
near the peak of the collimator transmission.  We believe that we can treat
Sco X-1 as a constant source for these purposes;  the presence of a nearly
stationary 6 Hz QPO
at the time of these observations indicates that Sco X-1 was in the normal
branch with a momentarily stable mass accretion and luminosity.

\begin{figure}
\psfig{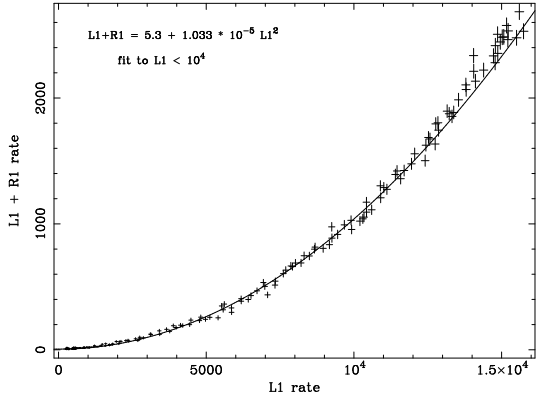}
\caption{L1+R1 rate fit as a quadratic function of L1 rate.  Both rates have been
corrected for ADC deadtime.  The fit is restricted to L1 $\le 10^4$ but provides
a reasonable description of the data at all observed rates.}
\label{fig4}
\end{figure}

Figure \ref{fig4} shows the same rate data, plotted against each other.
The rate data have been corrected for the non paralyzable deadtime associated
with the analog to digital conversion (and which must account for all events
in this detector, not just the two rates that are plotted).  Also shown
is the best fit which models the L1+R1 rate as a constant plus a term proportional
to the square of the L1 rate.  The fit considers only points with a corrected
L1 rate $\le 10^4$ count/sec, although the figure shows the fit extrapolated to the highest
observed rates.  The constant represents a background term, while the quadratic
coefficient represents the average time window during which a coincidence can be
recorded.  The average is over pulse height as we have ignored all energy information
while creating this correlation.  Identifying the L1 and R1 rates as the same
(due to their identical construction), we expect that the quadratic term
represents twice the coincidence window.  The chance of an L1+R1 event must
be equal to the L1 rate times the coincidence window time the R1 rate, and we must
multiply by two to account for R1+L1 events, which are indistinguishable to the EDS.
In this manner we identify the L1 coincidence window $\delta t_{1}$ as $5.02
\mus $.  

We use similar fits to estimate two additional time windows:  $\delta t_{2},
\delta t_{p}$ which are the coincidence windows associated
with the second or third xenon layers and the propane layer,  and a
window that describes the chance of getting an unflagged event $\delta t_{0}$.
Unflagged events occur for events with certain separations on the same signal
chains \cite{Jah96}.  The process is very slightly more complicated
when we fit Vp+L1 or Vp+R1 vs the L1 rate.  Here we assume that the window
for having an L1 event followed by a Vp event is known:  the window for
a Vp event followed by an L1 event is then the quadratic coefficient
minus $\delta t_{1}$.

The entire process was repeated using the 2 second readouts and a number
of bursts from J1744-28 (96Jan27).  Because the incident spectrum is quite different,
the derived coincidence windows are different, particularly on the first
layer.  Table 1 contains all of the derived windows.

\begin{table}
\setlength{\tabcolsep}{1.5pc}
\newlength{\digitwidth} \settowidth{\digitwidth}{\rm 0}
\catcode`?=\active \def?{\kern\digitwidth}

\label{tab:table1}
\caption{Coincidence timing windows}
\begin{tabular}{lrr}
\hline
	    & Sco X-1  &  J1744-28             \\
\cline{2-3}
	    & \multicolumn{2}{c} {$\mus$} \\
\hline
$\delta t_0$  &  5.0  &  6.5 \\
$\delta t_p$  &  4.5  &  3.2 \\
$\delta t_2$  &  9.2  &  9.5 \\
$\delta t_0$  &  1.8  &  4.2 \\
\hline
\end{tabular}
\end{table}

Two additional pieces of information are needed to infer incident rates
from observed coincidence rates.  First, the nonparalyzable deadtime associated with
the analog to digital conversion is taken to be $9 \mus$ \cite{Jah96}.
This value is independent of pulse height, and can be used for all
sources.  Second, we need the ratio of source related events which interact
in each layer of the detector, which is obviously dependent on the
spectrum.  For J1744-28 we rely on the fact that the bursts apparently have
the same spectrum as the persistent emission \cite{Giles96}, and use
the persistent emission to derive the ratios L1:L2:L3:VP to be
1000:150:75:180.   (The rate on R1 is assumed equal to that on L1, and
similarly for the other xenon layers).  For Sco X-1, we measure the
relative incident ratios from the scanning data obtained relatively far
from the peak and obtain 1000:67:21:440 (confirming our earlier statement 
that Sco X-1 has a much softer spectrum than J1744-28).

\subsection{Predicting the Coincidence Rate} 

Our convention is that ``incident rate" means the incident rate on the xenon
layers.  We can calculate the corresponding number of events observed in
the propane layer.  For a given incident rate, we calculate the incident
rate on each signal chain $R_j$ where the index $j$ runs from 1 to 7 and
corresponds to L1, R1, L2, R2, L3, R3, and VP.  The observed rate of
2 lower level discriminator events, $R_{2LLD}$ observed is given by
\begin{equation}
R_{2LLD} = A {\Sigma}_j {\Sigma}_{i \ne j} R_j \delta t_j R_i
\end{equation}
where $\delta t_j$ is defined as $\delta t_1$ for $j = 1,2$, as
$\delta t_2$ for $ j = 3,4,5,6 $, and as $\delta t_p $ for $j = 7$.
$A$  accounts for deadtime due to the non-paralyzable ADC process and is
calculated as $r_{in} / (1 + r_{in}t_d)$.
Since these are predicted rates, we have complete knowledge about the assumed
incident rate $r_{in}$.
The observed rate of unflagged events, $R_{noflag}$, and the observed rate of triply 
flagged events, $R_{3LLD}$, are similarly calculated:
\begin{equation}
R_{noflag} = A {\Sigma}_j R_j \delta t_0 R_j
\end{equation}
\begin{equation}
R_{3LLD} = A {\Sigma}_j {\Sigma}_{i \ne j} {\Sigma}_{k \ne i,j} R_j \delta t_j R_i R_k
\end{equation}
In order to predict the Remaining Count rate as recorded in the Standard 1 data,
all three terms must be added (for the bursts from J1744-28, the four fold
coincidence rate shows a detectable rise associated with the burst, but it
is not a significant contributor to the remaining rate).  
In all cases, $A$
must be evaluated considering all events (in other words, propane events
always contribute to the ADC deadtime).

\begin{figure}
\psfig{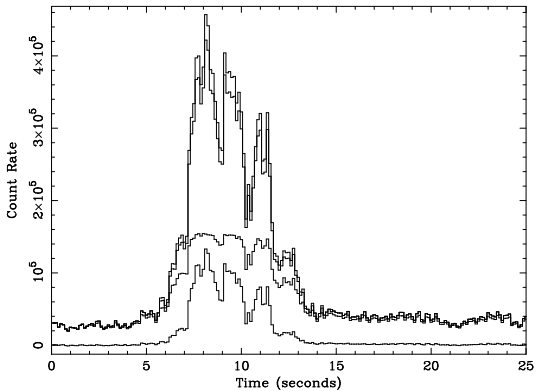}
\caption{Good count rate, Coincident count rate, and two different estimates
of the incident count rate.  See text for details.  The lower estimate of
the corrected rate is our best estimate.}
\label{fig5}
\end{figure}

Figure \ref{fig5} shows the same data as figure \ref{fig1}, along with the
corrected count rate.  We corrected the count rate twice, once using the
coefficients derived from J1744-28 and once using the coefficients derived
from Sco X-1.  In both cases we assume the relative count rate distribution between
the layers appropriate for J1744-28.  The higher correction, which peaks
at about $4.6 \times 10^5$ count/sec is associated with the coefficients for
Sco X-1, while the curve that peaks near $4.2 \times 10^5$ count/sec was
derived using the J1744-28 coefficients.  That the two curves differ by only
$10 \%$ demonstrates the expected robustness of this technique if one or the
other set of coefficients is used for a different bright source such as the
Soft Gamma Repeater SGR 1806-20.  It is easy to see why the Sco X-1 coefficients
result in a higher inferred incident rate.  Since we use the coincidence rate
to infer the incident rate, since the count rate on the L1 and R1 layers is
the most important contributor, and since the coincidence window derived from
J1744-28 data is {\it longer} than that derived from the Sco X-1 data, we
require a slightly lower incident rate to create the same double event rate.
While $10\%$ probably is a good estimate of the net uncertainty, we note that
we have reconstructed this estimate in the presence of $70 \%$ deadtime by
traditional definitions!  Our estimate is similar to an independent estimate
\cite{Fox97} which derives relationships between the various coincidence rates
and the good event rate without a detailed accounting for the different
coincidence windows on different layers.  While that model
produces a similar result, and is therefore of equal phenomenological
validity, they derive the value of the ADC deadtime, and consistently get a value
which is larger than the measured $9 \mus$.

\section{BURST HISTORY OF J1744-28}

\begin{figure}
\psfig{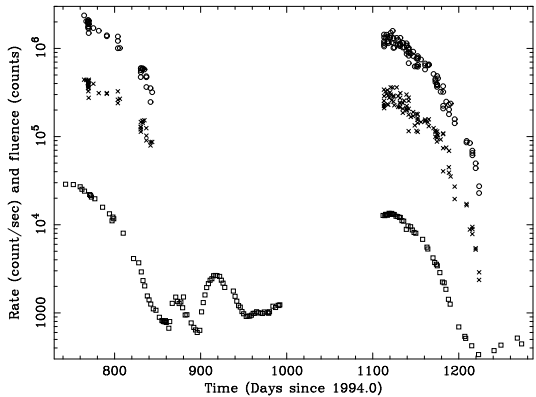}
\caption{Persistent flux (squares), peak burst flux (crosses), and burst
fluence (circles) for burst observed during observations of J1744-28.  Several
bursts defy the general trend;  these are largely Type I bursts from 
other sources in the field of view ([8]).}
\label{fig6}
\end{figure}

Finally we apply the same deadtime correction discussed above to the 
data base of J1744-28 observations.  RXTE has observed J1744-28 regularly
through out the first two years of the mission (\cite{Giles96} \cite{Stark98}).
Figure \ref{fig6} shows the persistent flux,
the peak flux in each burst, and the burst fluences.  We have not included
any bursts from May or June 1996;  bursts at this time are either Type I X-ray
bursts from another source within the field of view \cite{Stroh97a} or much fainter and
more frequent bursts which are part of a different ``rumbly" phenomenology \cite{Stark98}.
The bright bursts occur
only during the periods when the persistent flux is quite bright (above $\sim
1000$ count/sec).  However, we have observed bursts over a range of persistent
intensities that span one and a half orders of magnitude, and the striking
feature of fig \ref{fig6} is that the ratio of peak flux to persistent
flux, and the ratio of burst fluence to persistent flux, remain relatively constant.

Throughout the periods of bright bursting emission, the persistent emission
can be adequately described by a hard power law modified by galactic absorption
and a high energy exponential cutoff.  (There is also evidence for an Iron line
but this does not have a large effect on the derived flux).  This simple model
provides the same estimated flux as the continuous model described by \cite{biff98}.
We find that the 2-10 keV flux is $1.81 \times 10^{-12} \esc2$ per count/sec
(counts/sec measured in all PCA channels);  the 2-40 keV flux is $4.95 \times
10^{-12} \esc2$ per count/sec; the 2-100 keV flux is $5.40 \times 10^{-12} \esc2$
per count/sec.  Taking the highest estimated peak fluxes for bursts in the first 
outburst,
with peak rates of $4.4 \times 10^5$ count/sec,
we derive a peak flux of $2.4 \times 10^{-6} \esc2$.  The corresponding
peak luminosity is $1.8 \times 10^{40} d_8^2 \es$ where $d_8$ is the distance
to J1744-28 normalized to 8 kpc.  This is about 100 times the Eddington luminosity
given typical values for a neutron star mass and radius.  Given the nearly constant
ratio of burst fluencce to persistent flux, we conclude that the ratio $\alpha$
of persistent flux to time averaged burst flux must be approximately constant
(to the extent that the average interval between bursts is constant).  We
previously reported that $\alpha \sim 34$ in January 1997 \cite{Stroh97} and
take this as a representative value for both major outbursts.

None of the above conclusions could be reliably drawn from the good counting
rate alone, but require the extra information contained in the coincidence rate.
The most important conclsuion may therefore be our recommendation to future users
of the RXTE satellite who may be observing extremely bright sources:  use modes
which include some coincidence information at the highest timescales of interest.
On a practical level, extremely bright sources may be thought of as those
which produce $\ge 15,000$ count/sec/PCU with associated deadtimes of $\ge 15\%$.
An example of the value of this approach for observations of Sco X-1 is given
in \cite{vdk96}.

Comparisons to
the Crab nebula are often ambiguous due to the different spectral shapes.  However,
limiting the comparison to total count rate (13,000 count/sec for the Crab),
we find that the bursts reached 34 Crab.  Put another way, we can recast our
subtitle (using ``{\rm !}" for ``{\it i}"):  {\it  A grand one man band with major lighting:
stops at 17 ${4} \over {\rm two}$ Crab !!}

\vspace{0.5cm}
We thank the organizing committee and all participants for a stimulating workshop,
and Prof. Bignami for challenging us to present our result anagramatically.

\end{document}